\documentclass[12pt]{article}
\def\be{\beta}
\def\ga{\gamma}
\def\de{\delta}
\def\ep{\epsilon}

\def\ka{\kappa}
\def\la{\lambda}

\def\si{\sigma}

\def\ps{\psi}
\def\om{\omega}
\def\Ga{\Gamma}

\def\mn{{\mu\nu}}
\def\cl{{\mathcal L}}
\def\fr#1#2{{{#1} \over {#2}}}

\def\half{{\textstyle{1\over 2}}}
\def\frac#1#2{{\textstyle{{#1}\over {#2}}}}

\def\lsim{\mathrel{\rlap{\lower4pt\hbox{\hskip1pt$\sim$}}
    \raise1pt\hbox{$<$}}}
\def\gsim{\mathrel{\rlap{\lower4pt\hbox{\hskip1pt$\sim$}}
    \raise1pt\hbox{$>$}}}
\def\sqr#1#2{{\vcenter{\vbox{\hrule height.#2pt
         \hbox{\vrule width.#2pt height#1pt \kern#1pt
         \vrule width.#2pt}
         \hrule height.#2pt}}}}

\def\etal {{\it et al.}}
\newcommand{\beq}{\begin{equation}}
\newcommand{\eeq}{\end{equation}}
\newcommand{\bea}{\begin{eqnarray}}
\newcommand{\eea}{\end{eqnarray}}
\newcommand{\rf}[1]{(\ref{#1})}

\begin{document}

\title{Tests of Lorentz
and CPT Invariance in Space
\footnote{Contribution to the proceedings of the
2003 NASA/JPL Workshop on Fundamental
Physics in Space,
Oxnard, CA, April 2003.}
}
\author{Matthew Mewes\\
  \it Physics Department, Indiana University,\\
  \it Bloomington, IN 47405, U.S.A.}
\date{}
\maketitle

\begin{abstract}
I give a brief overview of recent work
concerning possible signals of Lorentz violation
in sensitive clock-based experiments in space.
The systems under consideration include atomic clocks
and electromagnetic resonators of the type planned
for flight on the International Space Station.
\end{abstract}

\section{Introduction}

In this contribution to the proceedings
of the 2003 NASA/JPL Workshop on Fundamental
Physics in Space, I review recent work
aimed at understanding possible tests of
Lorentz and CPT symmetries 
in experiments mounted on space platforms
such as the International Space Station (ISS)
\cite{ISS}.

A realistic description of nature
at the Planck scale remains a major
goal of theoretical physics.
A direct experimental search for Planck-scale
effects does not seem feasible using current technology.
However, it has been shown that
Planck-scale theories may lead to
small violations in fundamental
symmetries such as Lorentz and CPT covariance
in the low-energy effective theory \cite{kps}.
Such violations might arise out
of the nonlocal properties
of string theory.
Lorentz and CPT symmetries have also been
studied in the context of
noncommuting geometries \cite{ncqed}
and supersymmetry \cite{bek}.

Lorentz transformations are in general
comprised of rotations and boosts.
CPT is the combination of the discrete
transformations charge conjugation C,
space inversion P and time reversal T.
There is a general result known as the
CPT theorem which states that
a Lorentz-covariant theory is also
covariant under the combined
transformation CPT
\cite{owg}.

In recent years,
a number of sensitive experiments 
have tested Lorentz and CPT to
unprecedented levels \cite{cpt01}.
The increased activity in the field
has been motivated in part by
the development of a general
Lorentz- and CPT-violating
Standard-Model Extension (SME)
\cite{ck}.
The SME has provided a theoretical framework
for many tests of Lorentz and CPT covariance
including experiments involving
atomic systems \cite{ccexpt,lh,db,dp,kla,spaceexpt1,spaceexpt2},
photons \cite{miccavexpt,optcavexpt,photonexpt,photonth,km},
hadrons \cite{hadronexpt,hadronth},
muons \cite{muons},
and electrons \cite{eexpt,eexpt2}.

One particularly sensitive class
of experiments involves extremely
precise clocks and resonators.
A number of experiments of this type
are under development to test relativity
principles on the ISS.
These include
the atomic-clock based experiments
ACES \cite{aces},
PARCS \cite{parcs},
RACE \cite{race}
and a resonant-cavity experiment,
SUMO \cite{sumo}.
Some of the best constraints on Lorentz
and CPT violation have been achieved
in Earth-based atomic-clock experiments
\cite{ccexpt,lh,db,dp,kla}.
Recently, similar techniques have
been used in Earth-based experiments involving
superconducting microwave cavities
\cite{miccavexpt}
and cryogenically cooled optical cavities
\cite{optcavexpt}
that probed previously
untested regions of coefficient space.
The basic principle behind all these
experiments is to search for
variations in frequencies of
resonant systems as the Earth rotates.
The space-based versions will look for
variations as the satellite orbits the Earth.

Here, I review recent theoretical
studies concerning the effects of Lorentz
and CPT violation on atomic clocks
\cite{spaceexpt1,spaceexpt2}
and resonant cavities \cite{km}
aboard orbiting platforms such as the ISS.
A brief discussion of the SME
and the QED limit can be
found in Section \ref{qed}.
A general discussion of the types
of signals one expects from Lorentz
violation are described in Section \ref{sig}.
Some results in atomic clocks and
in resonant cavities are given
Sections \ref{ac} and \ref{rc}.
Some advantages of space-based
experiments are described
in Section \ref{adv}.

\section{Lorentz-Violating QED}\label{qed}

The purpose of the SME is the characterization
of all possible types of Lorentz violation
in a single local relativistic quantum field theory.
Under mild assumptions,
one finds that the form of the theory is
restricted to the usual Standard-Model
lagrangian supplemented by terms that consist
of Standard-Model field combinations multiplied
by small constant coefficients \cite{ck}.
Each term must form a scalar under
Lorentz transformations of the observer
so that coordinate invariance is satisfied.
Often one restricts attention to renormalizable terms.
However, the nonrenormalizable sector is
known to be important at very high energies
\cite{kle}.

The QED limit of the SME serves as a
toy-model example of this general framework.
It also has physical significance since
many systems are accurately represented
by this limit.
The renormalizable sector of the QED
extension is given by the lagrangian
\bea
{\cl} &=& \frac{1}{2} i \bar{\ps} \Ga^\nu
\stackrel{\leftrightarrow}{D_\nu} \ps  -  \bar{\ps}
M \ps  - \frac{1}{4} F^{\mu\nu} F_{\mu\nu} 
\nonumber \\ & & 
- \frac {1}{4}
(k_F)_{\ka\la\mu\nu} F^{\ka\la} F^{\mu\nu} + \frac{1}{2}
(k_{AF})^{\ka} \ep_{\ka\la\mu\nu} A^\la F^{\mu\nu},
\label{lag}
\eea
where $D_\mu$ is the usual covariant derivative and
\bea
\Ga^{\nu} &=&
\ga^{\nu} + c^\mn \ga_\mu + d^\mn \ga_5
\ga_\mu + e^\nu + i f^\nu \ga_5 + \half g^{\la\mn}
\si_{\la\mu}, \\
M &=& m + a_\mu \ga^\mu + b_\mu \ga_5 \ga^\mu +
\half H_{\mn} \si^{\mn}.
\label{coeff}
\eea
The small coefficients
$c^\mn$, $d^\mn$, $H_{\mn}$ and $(k_F)_{\ka\la\mu\nu}$
introduce Lorentz violation and are CPT even.
Meanwhile, $e^\nu$, $f^\nu$, $g^{\la\mn}$,
$a_\mu$, $b_\mu$ and $(k_{AF})^{\ka}$
are Lorentz violating and CPT odd.
Note that taking these coefficients to zero
yields the usual QED.

The experiments considered in this work
search for frequency shifts due to the
above coefficients.
For atomic clocks, the frequency is
typically determined by Zeeman transitions.
The presence of Lorentz and CPT
violation results in small shifts in
these transitions that
depend on the coefficients
in the modified QED associated with
each of the particle species:
protons, neutrons and electrons.
These coefficients are denoted
$a^w_\mu$, $b^w_\mu$, $c^w_\mn$, $d^w_\mn$,
$e^w_\nu$, $f^w_\nu$, $g^w_{\la\mn}$, $H^w_{\mn}$,
where the $w=p,n,e$ labels the species
\cite{kla}.

In practice, only certain combinations
of coefficients appear.
These are commonly 
denoted by tilde coefficients
$\tilde b^w_3$, $\tilde c^w_q$,
$\tilde d^w_3$, $\tilde g^w_d$,
$\tilde g^w_q$,
where I have assumed that the
quantization axis is along the 3 direction.
As an example of the relationship
between the tilde coefficients and those
in Eq.\ \rf{coeff} consider $b^e_3$.
It is related to the coefficients in the
QED for electrons by the expression
$\tilde b^e_3=b^e_3-m_ed^e_{30}+m_eg^e_{120}-H^e_{12}$.
The subscript 3 refers to
the quantization axis in the laboratory
which in this example was chosen
to be in the 3 direction.
The subscripts $d$ and $q$ refer to
the dipole and quadrupole nature
of those terms.

A similar tilde decomposition is
useful in the photon sector.
The presence of Lorentz violation
leads to similar shifts in
the resonant frequencies of cavities.
When calculating these shifts
it is useful to work with
the usual electric and magnetic fields.
In terms $\vec E$ and $\vec B$,
the $(k_F)_{\ka\la\mu\nu}$ term in the
lagrangian \rf{lag} may be written
\bea
-\frac14(k_F)_{\ka\la\mu\nu} F^{\ka\la} F^{\mu\nu}
&=&\half\tilde\ka_{\rm tr}(\vec E^2+\vec B^2)
+\half \vec E\cdot(\tilde\ka_{e+}+\tilde\ka_{e-})\cdot\vec E
\nonumber \\
&&-\half\vec B\cdot(\tilde\ka_{e+}-\tilde\ka_{e-})\cdot\vec B 
+\vec E\cdot(\tilde\ka_{o+}+\tilde\ka_{o-})\cdot\vec B .
\label{kf}
\eea 
The subscripts $e$, $o$ and $\rm tr$
refer to their O(3) properties.
The coefficients
$\tilde\ka_{\rm tr}$,
$\tilde\ka_{e+}$ and
$\tilde\ka_{e-}$ are parity even
while
$\tilde\ka_{o+}$ and
$\tilde\ka_{o-}$ are parity odd.
The single coefficient
$\tilde\ka_{\rm tr}$
is rotationally invariant
while the others are $3\times3$
traceless matrices that violate
rotational symmetry.

The above decomposition is motivated
by constraints on birefringence of light
originating from very distant galaxies.
Nonozero coefficients
$\tilde\ka_{e+}$ and $\tilde\ka_{o-}$
cause birefringence in light
as it traverses empty space
resulting in a well defined
energy dependence in its polarization.
Spectropolarimetric observations of
light emitted from distant radio
galaxies places a limit on this effect
and leads to constraints
on the order of $\sim 10^{-32}$ on 
$\tilde\ka_{e+}$ and $\tilde\ka_{o-}$
\cite{km}.

\section{Signatures of Lorentz Violation}\label{sig}

In the event of appreciable Lorentz violation,
we would expect experiments to
depend on their orientation
since rotations are a subgroup
of Lorentz transformations.
The Lorentz group also contains
boosts which implies we would
expect velocity dependence as well.
A common approach in tests of
Lorentz covariance is to search
for these types of dependences
by looking for variations in
some observable as the Earth rotates
and orbits the Sun.
The rotation of the Earth causes
changes in the orientation of
the apparatus,
while the orbital motion
results in changes in velocity.
Note that boost effects resulting
from the change in velocity are
typically suppressed by a factor of
$\be_{\oplus}\sim10^{-4}$, the velocity
of the Earth around the Sun.

To understand how the orientation and
velocity dependence is quantified,
we must define at least two
frames of reference.
The first is the laboratory frame
with coordinates $(0,1,2,3)$.\footnote{
A standard set of frames for
Earth-based and satellite-based experiments
is defined in Ref.\ \cite{km}.}
The clock or cavity is at rest
in this frame which simplifies calculations.
In these experiments the Lorentz violation
typically leads to frequency shifts that are
linear in the tilde coefficients
discussed in the previous section.
However, these tilde coefficients are
not necessarily constant since
they are associated with the $(0,1,2,3)$ frame
which is not inertial.

To express the frequency shifts in terms
of constant coefficients we must choose
an inertial frame of reference.
The conventional choice is a standard
Sun-centered celestial equatorial frame
with coordinates $(T,X,Y,Z)$.
This frame may be considered inertial
for all practical purposes and provides
a common set of coefficients which
all experiments can refer to.
We can relate the coefficients in the
$(0,1,2,3)$ frame to those in the 
$(T,X,Y,Z)$ frame by a Lorentz transformation
which is time-dependent since the laboratory
frame is in constant motion.
For Earth-based experiments this typically
introduces a periodic variation at the
Earth's rotation rate
$\om_\oplus\simeq 2\pi/$(23 h, 56 min.)
and at $2\om_\oplus$,
providing a signal for Lorentz violation.
Similar variations at the orbital frequency
$\om_s\simeq 92$ min.\ and $2\om_s$
occur in experiments aboard the ISS.

\subsection{Atomic Clocks in Space}\label{ac}

As an example, here I briefly discuss
how atomic-clock experiments on the ISS
could be used to search for Lorentz violation.
For details I refer the reader to the
recent analyses found in Refs.\ 
\cite{spaceexpt1,spaceexpt2}.

A typical clock-comparison
experiment consists of two
co-located clocks using
different atomic species
or operating on different transitions.
Each species and
transition responds differently
to Lorentz violation.
If we compare the signals
from the two clocks we may be able
to detect a relative shift in their frequencies.
For simplicity, one clock could operate
on a transition that is known
to be insensitive to Lorentz violation
\cite{kla}.

Consider a clock at rest in the
ISS frame with its quantization
axis along the 3 direction.
In general, the frequency shift
depends on the combinations
$\tilde b^w_3$, $\tilde c^w_q$,
$\tilde d^w_3$, $\tilde g^w_d$,
$\tilde g^w_q$.
The instantaneous values of
these coefficients determine the frequency
of the clock at any point in the orbit.
Expressing these coefficients
in terms of Sun-frame coefficients
reveals time dependence not present
in the absence of Lorentz violation.
It is this time dependence that
provides a discernible signal
for violations in Lorentz and
CPT covariance.

The full expressions relating the
coefficients in each frame are rather lengthy.
However, to first order in small velocities,
they take the form \cite{spaceexpt2}:
\bea
\tilde b_3,\tilde d_3,\tilde g_d
&=&\cos\om_sT_s [\sim]
+\sin\om_sT_s [\sim]
\nonumber\\
&&+\be_s\cos2\om_sT_s [\sim]
+\be_s\sin2\om_sT_s [\sim]
+\be_s[\sim] ,\label{cc1}\\
\tilde c_q,\tilde g_q
&=&\be_s\cos\om_sT_s [\sim]
+\be_s\sin\om_sT_s [\sim]
\nonumber\\
&&+\cos2\om_sT_s [\sim]
+\sin2\om_sT_s [\sim]
+[\sim] ,\label{cc2}
\eea
where each $[\sim]$ indicates
a different linear combination of the
Sun-frame tilde coefficients
$\tilde b_T$, $\tilde b_X$,
$\tilde b_Y$, $\tilde b_Z$,
$\tilde g_T$, $\cdots$ .
The quantities $\om_s\simeq2\pi/$92 min.\ 
and $\be_s\simeq 10^{-5}$ are the
frequency and velocity of the ISS orbit
and $T_s$ is the time with an appropriately
chosen zero.
Note that the $2\om_s$ variations
in the vector and dipole coefficients
and the $\om_s$ variations in the
quadrupole terms are suppressed by $\be_s$.

\subsection{Resonant Cavities in Space}\label{rc}

Also slated to fly aboard the ISS
is the SUMO experiment \cite{sumo}.
This experiment utilizes superconducting
microwave oscillators.
The frequencies of resonant cavities
are also shifted by Lorentz violation.
However, they are sensitive to the photon
sector of the QED extension.
A detailed analysis of the effects of
the $\tilde\ka$ coefficients on the
resonant frequencies of cavities can
be found in Ref.\ \cite{km}.
The results relevant to SUMO are
summarized below.

The cavities used in SUMO are
cylindrical with circular cross section
and operate in the fundamental TM$_{010}$ mode.
Working in a frame where the symmetry
axis coincides with 3 axis,
a perturbative calculation finds that
the frequency shift is linear in the
coefficient combinations
$(3\tilde\ka_{e+}+\tilde\ka_{e-})^{33}$
and $\tilde\ka_{\rm tr}$.
The frequency shift is easily generalized to
a cavity that is at rest in the laboratory
but arbitrarily oriented with its
symmetry axis denote by a unit vector $\hat N$.
The result is the fractional-frequency shift
\beq
\fr{\de\nu}{\nu} =
-\frac14\hat N^j\hat N^k
(3\tilde\ka_{e+}+\tilde\ka_{e-})^{jk}
-\tilde\ka_{\rm tr}\ ,
\label{dnu1}
\eeq
where the indices sum over 
laboratory-frame coordinates, $j,k=1,2,3$.
This expression is valid in any laboratory
frame at rest with respect to the cavity.

In order to fully understand
the effects of Lorentz violation
on a cavity in orbit,
we must transform the coefficients
to the Sun-centered frame.
To first order in the boost velocity,
the answer can be written
\beq
\fr{\de\nu}{\nu} =
-\frac14\hat N^j\hat 
 N^kR^{jJ}R^{kK}(\tilde\ka_{e'})^{JK}
-\half (\de^{jk}+\hat N^j\hat N^k)
R^{jJ}R^{kK}\ep^{JPQ}\be^Q (\tilde\ka_{o'})^{KP}
-\tilde\ka_{\rm tr}\ ,
\label{dnu2}
\eeq
where for convenience we define
\beq
(\tilde\ka_{e'})^{JK}=3(\tilde\ka_{e+})^{JK}
+(\tilde\ka_{e-})^{JK}\ ,\quad
(\tilde\ka_{o'})^{JK}=3(\tilde\ka_{o-})^{JK}
+(\tilde\ka_{o+})^{JK} .
\eeq
The uppercase indices represent
the Sun-centered coordinates, $J,K=X,Y,Z$.
The matrix $R$ is the rotation
between the two frames and
$\be$ is the velocity of the
laboratory in the Sun frame.
Inserting the explicit time-dependent
expressions $R$ and $\be$
leads to periodic variations similar
to the atomic-clock case.

A number of different experiments are possible.
For example, a cavity could be
compared to an atomic clock.
The clock could be used as reference
by choosing a transition that is
insensitive to Lorentz violation.
This setup would only be sensitive to
violations in the photon sector.
In contrast,
operating the clock on a transition
sensitive to Lorentz violation would
provide sensitivity to combinations
of photon and fermion coefficients.

It is also possible to construct
cavities that are insensitive to
given tilde coefficients.
For example, geometries exist
that support modes that
are insensitive to $\tilde\ka_{e-}$.
With the constraints from birefringence,
this leaves only the $\be$ suppressed
variations due to  $\tilde\ka_{o+}$.
Therefore, cavities might serve as
reference frequencies for atomic clocks.

Traditionally, two cavities oriented
at right angles are used in tests
of relativity.
This method could also be implemented in
space-based experiments.
In two-cavity experiments,
the quantity of interest is normally
the beat frequency obtained by
combining their signals.
On the ISS, this will take the form
\beq
\fr{\nu_{beat}}{\nu} \equiv 
\fr {\de\nu_1} {\nu} - \fr{\de\nu_2}{\nu}
= {\cal A}_s\sin\om_sT_s +{\cal A}_c\cos\om_sT_s 
+{\cal B}_s\sin2\om_sT_s+{\cal B}_c\cos2\om_sT_s + {\cal C} ,
\label{dnu3}
\eeq
where the amplitudes
${\cal A}_s$, ${\cal A}_c$,
${\cal B}_s$, and ${\cal B}_c$
are linear combinations
of the tilde coefficients.
These are typically rather cumbersome
\cite{km} but depend
on the orientation of the cavity pair
in the laboratory and
on the orientation of the orbital plane 
with respect to the Sun-centered frame.

It can be shown that orienting
a cavity with $\hat N$
in the orbital plane
maximizes the sensitivity to
the second harmonics, at leading
order in $\be$ and that
orienting a cavity so that
$\hat N$ is $45^\circ$ out of the
plane maximizes sensitivity to
the first harmonics.
Therefore, a sensible configuration might 
have one cavity in the
orbital plane and one $45^\circ$
out of it.

\section{Advantages of Space-Based Experiments}\label{adv}

There are several advantages
to space-based experiments
over their ground-based counterparts.
A major advantage stems from
the relatively short orbital
period of the ISS.
In Earth-based experiments,
the relevant period is one sidereal day.
Comparing this to the 92 min.\ 
period of the ISS orbit implies
that an experiment on the ISS
could acquire a comparable dataset
in approximately one-sixteenth the time.

Another advantage arises from the
properties of the ISS orbital plane.
For fixed Earth-based experiments,
there are combinations of coefficients
such as $\tilde b_Z$ and
$(\tilde\ka_{e-})^{ZZ}$ that do not
contribute to sidereal variations and
are therefore unobservable.\footnote{
Coefficients of this type
can be accessed with the use of
a turntable as in Ref.\ \cite{eexpt2}.}
This is due to the constancy of the
Earth's rotational axis which
is fixed and points in the $\hat Z$ direction.
The analogous direction in the
case of the ISS is given by its orbital axis.
However, this axis precesses about
the $\hat Z$ axis at an
angle of approximately $52^\circ$,
implying that there is no analogous
set of inaccessible coefficients.

One last major advantage
is due to $\be$ suppressed
terms like those that appear
in Eqs.\ \rf{cc1} and \rf{cc2}.
Note that similar $\be_\oplus$ and $\be_s$
suppressed terms appear in the
$[\sim]$ combinations of
Eqs.\ \rf{cc1} and \rf{cc2}
and in the amplitudes of Eq.\ \rf{dnu3}.
These terms are due to the
changing velocity of the ISS
in the Sun frame and
introduce new time dependences
and sensitivities to
coefficient combinations that
do not appear when
considering rotational effects alone.
Analogous terms do arise in
Earth-based experiments.
However, the terms that introduce
new time dependences are suppressed by
the smaller laboratory velocity
$\be_L\lsim 1.5\times10^{-6}\ll\be_s$.

\section{Summary and Discussion}

\begin{table}
\begin{center}
\begin{tabular}{cccc}
\hline
\hline
Coefficient & Birefringence & Microwave & Optical \\
\hline
$(\tilde\ka_{e+})^{JK}$ & -32 & * & * \\
$(\tilde\ka_{o-})^{JK}$ & -32 & * & * \\
$(\tilde\ka_{e-})^{XX}
-(\tilde\ka_{e-})^{YY}$ & n/a & -13 & -15\\
$(\tilde\ka_{e-})^{ZZ}$ & n/a & - & - \\
$(\tilde\ka_{e-})^{XY}$,
$(\tilde\ka_{e-})^{XZ}$,
$(\tilde\ka_{e-})^{YZ}$ & n/a & -13 & -15 \\
$(\tilde\ka_{o+})^{XY}$,
$(\tilde\ka_{o+})^{XZ}$,
$(\tilde\ka_{o+})^{YZ}$ & n/a & -9 & -11 \\
$\tilde\ka_{\rm tr}$ & n/a & - & -\\
\hline
\hline
\end{tabular}
\caption{\label{phtab}
Existing bounds for
cosmological birefringence \cite{km},
microwave cavities \cite{miccavexpt} and
optical cavities \cite{optcavexpt}.
A star indicates that constraints probably exist.
However, to date, no analysis has
included these coefficients.}
\end{center}
\end{table}
\begin{table}
\begin{center}
\begin{tabular}{cccc} 
\hline \hline 
Coefficient & Proton & Neutron & Electron \\
\hline
$\tilde b_X$, $\tilde b_Y$   & -27[-27]& [-31]   & -27[-29]  \\
$\tilde b_Z$                 & -27     & -       & -27[-28]   \\
$\tilde b_T$                 & -23     & -       & -23     \\
$\tilde g_T$                 & -23     & -       & -23     \\
$\tilde H_{JT}$              & -23     & -       & -23     \\
$\tilde d_\pm$               & -23     & -       & -23     \\
$\tilde d_Q$                 & -23     & -       & -23     \\
$\tilde d_{JK}$              & -23     & -       & -23     \\
$\tilde d_X$, $\tilde d_Y$        & -25[-25]& [-29]   & -22[-22]  \\
$\tilde d_Z$                 & -25     & -       & -22     \\
$\tilde g_{DX}$,$\tilde g_{DY}$   & -25[-25]& [-29]   & -22[-22]  \\
$\tilde g_{DZ}$              & -25     & -       & -22     \\
$\tilde g_{JK}$              & -21     & -       & -18     \\
$\tilde g_c$                 & -23     & -       & -23     \\
$\tilde c_{TJ}$              & -20     & -       & -       \\
$\tilde c_-$                 & -25     & [-27]     & -       \\
$\tilde c_Q$                 & -25     & -       & -       \\
$\tilde c_X$, $\tilde c_Y$        & -25     & [-25]     & -       \\
$\tilde c_Z$                 & -25     & [-27]     & -       \\
$\tilde c_{TJ}$              & -21     & -       & -       \\
$\tilde g_-$                 & $\star$[$\star$] & [$\star$] & -       \\
$\tilde g_Q$                 & $\star$ & -       & -       \\
$\tilde g_{TX}$, $\tilde g_{TY}$  & $\star$[$\star$] & [$\star$] & -       \\
$\tilde g_{TZ}$              & $\star$[$\star$] & [$\star$] & -       \\ 
\hline \hline 
\end{tabular}
\caption{\label{cctab}
Estimated sensitivity 
to tilde coefficients
for ISS experiments with
$^{133}$Cs and $^{87}$Rb clocks
taken from Ref.\ \cite{spaceexpt2}.
Existing bounds 
\cite{ccexpt,lh,db,dp,eexpt2}
are shown in brackets.
A star indicates possible
sensitivity in realistic nuclear model.}
\end{center}
\end{table}

Table \ref{phtab} lists the approximate
base-10 logarithm of existing
constraints on Lorentz violation
in the photon sector.
Ground-based experiments involving
microwave \cite{miccavexpt}
and optical \cite{optcavexpt}
cavities have measured all
components of $\tilde\ka_{e-}$ and
$\tilde\ka_{o+}$ except
$(\tilde\ka_{e-})^{ZZ}$.
A space-based experiment could
immediately access the unconstrained
coefficient $(\tilde\ka_{e-})^{ZZ}$.
Improved sensitivities are also expected.
It has been estimated that SUMO may
be able to achieve sensitivity
at the level of $10^{-17}$
\cite{miccavexpt}.

The above discussion could can also
be applied to optical-cavity experiments.
Currently, the most precise measurements
of $\tilde\ka_{e-}$ and $\tilde\ka_{o+}$
are from an optical-cavity experiment
\cite{optcavexpt} and space-based
versions  such as those proposed for the 
OPTIS experiment \cite{optis}
could also yield interesting results.

Note that the rotationally
invariant component $\tilde\ka_{\rm tr}$
is also unconstrained.
This is because, at order $\be$,
it results in unobservable constant shifts.
However, it becomes important
at order  $\be^2$ and could
be accessed at interesting levels
in experiments involving larger boosts
or better sensitivity.

Table \ref{cctab} lists
the estimates given in Ref.\ \cite{spaceexpt2}
for the sensitivities of 
$^{133}$Cs and $^{87}$Rb
clocks on the ISS.
The brackets indicate measurements
from current ground-based experiments.
The table illustrates the main
advantage of space-based
clock-comparison experiments.
The additional freedom in the
motion of the ISS results
in access to a much larger portion
of the coefficient space.

Future clock-comparison experiments
in space will probe regions of coefficient
space difficult to access on Earth.
They will do it more quickly and perhaps
with better sensitivity than
their ground-based counterparts.


\begin{thebibliography}{99}

\bibitem{ISS}
See, for example,
\it The International Space Station Users' Guide, \rm
Release 2.0, NASA, 2000.

\bibitem{kps}
V.A.\ Kosteleck\'y and R.\ Potting,
Nucl.\ Phys.\ B {\bf 359}, 545 (1991);
Phys.\ Lett.\ B {\bf 381}, 89 (1996);
Phys.\ Rev.\ D {\bf 63}, 046007 (2001);
V.A.\ Kosteleck\'y and S.\ Samuel,
Phys.\ Rev.\ D {\bf 39}, 683 (1989);
Phys.\ Rev.\ Lett.\ {\bf 63}, 224 (1989);
Phys.\ Rev.\ D {\bf 40}, 1886 (1989).
V.A.\ Kosteleck\'y, M.\ Perry, and R.\ Potting,
Phys.\ Rev.\ Lett.\ {\bf 84}, 4541 (2000).

\bibitem{ncqed}
S.M.\ Carroll \etal,
Phys.\ Rev.\ Lett.\ {\bf 87}, 141601 (2001);
Z.\ Guralnik, R.\ Jackiw, S.Y.\ Pi, and A.P.\ Polychronakos,
Phys.\ Lett.\ B {\bf 517}, 450 (2001);
C.E.\ Carlson, C.D.\ Carone, and R.F.\ Lebed,
Phys.\ Lett.\ B {\bf 518}, 201 (2001);
A.\ Anisimov, T.\ Banks, M.\ Dine, and M.\ Graesser,
Phys.\ Rev.\ D {\bf 65}, 085032 (2002);
I.\ Mocioiu, M.\ Pospelov, and R.\ Roiban,
Phys.\ Rev.\ D {\bf 65}, 107702 (2002);
M.\ Chaichian, M.M.\ Sheikh-Jabbari, and A.\ Tureanu,
hep-th/0212259;
J.L.\ Hewett, F.J.\ Petriello, and T.G.\ Rizzo,
Phys.\ Rev.\ D {\bf 66}, 036001 (2002).

\bibitem{bek}
M.S.\ Berger and V.A.\ Kosteleck\'y,
Phys.\ Rev.\ D {\bf 65}, 091701(R) (2002).

\bibitem{owg}
O.W.\ Greenberg,
Phys.\ Rev.\ Lett.\ {\bf 89}, 231602 (2002);
hep-ph/0305276.

\bibitem{cpt01}
For recent overviews of various experimental 
and theoretical approaches to Lorentz and CPT violation,
see, for example,
V.A.\ Kosteleck\'y, ed.,
{\it CPT and Lorentz Symmetry II},
World Scientific, Singapore, 2002.

\bibitem{ck} 
D.\ Colladay and V.A.\ Kosteleck\'y,
Phys.\ Rev.\ D {\bf 55}, 6760 (1997);
{\bf 58}, 116002 (1998).

\bibitem{ccexpt}
V.W.\ Hughes, H.G.\ Robinson, and V.\ Beltran-Lopez,
Phys.\ Rev.\ Lett.\ {\bf 4} (1960) 342;
R.W.P.\ Drever,
Philos.\ Mag.\ {\bf 6} (1961) 683;
J.D.\ Prestage
{\it et al.},
Phys.\ Rev.\ Lett.\ {\bf 54} (1985) 2387;
S.K.\ Lamoreaux
{\it et al.},
Phys.\ Rev.\ A {\bf 39} (1989) 1082;
T.E.\ Chupp
{\it et al.},
Phys.\ Rev.\ Lett.\ {\bf 63} (1989) 1541.

\bibitem{lh}
C.J.\ Berglund
{\it et al.},
Phys.\ Rev.\ Lett.\ {\bf 75} (1995) 1879;
L.R.\ Hunter
{\it et al.},
in 
V.A.\ Kosteleck\'y, ed.,
\it CPT and Lorentz Symmetry, \rm
World Scientific, Singapore, 1999.

\bibitem{db}
D.\ Bear
{\it et al.},
Phys.\ Rev.\ Lett.\ {\bf 85}, 5038 (2000).

\bibitem{dp}
D.F.\ Phillips
{\it et al.},
Phys.\ Rev.\ D {\bf 63}, 111101 (2001);
M.A.\ Humphrey 
{\it et al.},
physics/0103068;
Phys.\ Rev.\ A {\bf 62}, 063405 (2000).

\bibitem{kla}
V.A.\ Kosteleck\'y and C.D.\ Lane,
Phys.\ Rev.\ D {\bf 60}, 116010 (1999);
J.\ Math.\ Phys.\ {\bf 40}, 6245 (1999).

\bibitem{spaceexpt1}
R.\ Bluhm \etal,
Phys.\ Rev.\ Lett.\ {\bf 88}, 090801 (2002).

\bibitem{spaceexpt2}
R.\ Bluhm \etal, hep-ph/0306190.

\bibitem{miccavexpt}
J.\ Lipa
{\it et al.},
Phys.\ Rev.\ Lett.\ {\bf 90}, 060403 (2003).

\bibitem{optcavexpt}
H.\ M\"uller
{\it et al.},
physics/0305117.

\bibitem{photonexpt}
S.M.\ Carroll, G.B.\ Field, and R.\ Jackiw, 
Phys. Rev. D {\bf 41}, 1231 (1990);
V.A.\ Kosteleck\'y and M.\ Mewes,
Phys.\ Rev.\ Lett.\ {\bf 87}, 251304 (2001).

\bibitem{photonth}
R.\ Jackiw and V.A.\ Kosteleck\'y,
Phys.\ Rev.\ Lett.\ {\bf 82}, 3572 (1999);
C.\ Adam and F.R.\ Klinkhamer,
Nucl.\ Phys.\ B {\bf 657}, 214 (2003);
H.\ M\"uller, C.\ Braxmaier, S.\ Herrmann, 
A.\ Peters, and C.\ L\"ammerzahl,
Phys. Rev. D {\bf 67}, 056006 (2003);
T.\ Jacobson, S.\ Liberati, and D.\ Mattingly,
hep-ph/0209264;
V.A.\ Kosteleck\'y, M.\ Perry, and R.\ Lehnert,
astro-ph/0212003;
V.A.\ Kosteleck\'y and A.G.M.\ Pickering,
Phys.\ Rev.\ Lett., in press (hep-ph/0212382);
R.\ Lehnert, gr-qc/0304013;
G.M.\ Shore, gr-qc/0304059.

\bibitem{km}
V.A.\ Kosteleck\'y and M.\ Mewes,
Phys.\ Rev.\ D {\bf 66}, 056005 (2002).

\bibitem{hadronexpt}
KTeV Collaboration,
H.\ Nguyen, in Ref.\ \cite{cpt01};
OPAL Collaboration,
R.\ Ackerstaff
{\it et al.},
Z.\ Phys.\ C {\bf 76}, 401 (1997);
DELPHI Collaboration,
M.\ Feindt
{\it et al.},
preprint DELPHI 97-98 CONF 80 (1997);
BELLE Collaboration,
K.\ Abe
{\it et al.},
Phys.\ Rev.\ Lett.\ {\bf 86}, 3228 (2001);
BaBar Collaboration,
B.\ Aubert
{\it et al.}, 
hep-ex/0303043;
FOCUS Collaboration,
J.M.\ Link 
{\it et al.}, 
Phys.\ Lett.\ B {\bf 556}, 7 (2003).

\bibitem{hadronth}
V.A.\ Kosteleck\'y and R.\ Potting,
Phys.\ Rev.\ D {\bf 51}, 3923 (1995);
D.\ Colladay and V.A.\ Kosteleck\'y,
Phys.\ Lett.\ B {\bf 344}, 259 (1995);
Phys.\ Rev.\ D {\bf 52}, 6224 (1995);
Phys.\ Lett.\ B {\bf 511}, 209 (2001);
V.A.\ Kosteleck\'y and R.\ Van Kooten,
Phys.\ Rev.\ D {\bf 54}, 5585 (1996);
O.\ Bertolami
{\it et al.},
Phys.\ Lett.\ B {\bf 395}, 178 (1997);
V.A.\ Kosteleck\'y,
Phys.\ Rev.\ Lett.\ {\bf 80}, 1818 (1998);
Phys.\ Rev.\ D {\bf 61}, 016002 (2000);
{\bf 64}, 076001 (2001);
N.\ Isgur \etal,
Phys.\ Lett.\ B {\bf 515}, 333 (2001).

\bibitem{muons} 
V.W.\ Hughes
{\it et al.},
Phys.\ Rev.\ Lett.\ {\bf 87}, 111804 (2001);
R.\ Bluhm \etal,
Phys.\ Rev.\ Lett.\ {\bf 84}, 1098 (2000).

\bibitem{eexpt}
H.\ Dehmelt 
{\it et al.},
Phys.\ Rev.\ Lett.\ {\bf 83}, 4694 (1999);
R.\ Mittleman 
{\it et al.},
Phys.\ Rev.\ Lett.\ {\bf 83}, 2116 (1999);
G.\ Gabrielse 
{\it et al.},
Phys.\ Rev.\ Lett.\ {\bf 82}, 3198 (1999);
R.\ Bluhm \etal,
Phys.\ Rev.\ Lett.\ {\bf 82}, 2254 (1999);
Phys.\ Rev.\ Lett.\ {\bf 79}, 1432 (1997);
Phys.\ Rev.\ D {\bf 57}, 3932 (1998).

\bibitem{eexpt2}
B.\ Heckel,
in Ref.\ \cite{cpt01};
L.-S.\ Hou, W.-T.\ Ni, and Y.-C.M.\ Li,
Phys.\ Rev.\ Lett.\ {\bf 90}, 201101 (2003);
R.\ Bluhm and V.A.\ Kosteleck\'y,
Phys.\ Rev.\ Lett.\ {\bf 84}, 1381 (2000).

\bibitem{aces}
P.\ Laurent
{\it et al.},
Eur.\ Phys.\ J.\ D {\bf 3} (1998) 201.

\bibitem{parcs}
N.\ Ashby,
in Ref.\ \cite{cpt01}.

\bibitem{race}
C.\ Fertig 
{\it et al.},
Proceedings of the Workshop on Fundamental Physics in Space,
Solvang, June 2000;
and in Ref.\ \cite{cpt01}.

\bibitem{sumo}
S.\ Buchman 
{\it et al.},
Adv.\ Space Res.\ {\bf 25}, 1251 (2000);
J.\ Nissen
{\it et al.},
in Ref.\ \cite{cpt01}.

\bibitem{kle}
V.A.\ Kosteleck\'y and R.\ Lehnert,
Phys.\ Rev.\ D {\bf 63}, 065008 (2001);
V.A.\ Kosteleck\'y, C.D.\ Lane, and A.G.M.\ Pickering,
Phys.\ Rev.\ D {\bf 65}, 056006 (2002).

\bibitem{optis}
C.\ L\"ammerzahl
{\it el al.},
Class.\ Quant.\ Gravity {\bf 18}, 2499 (2001).

\end{thebibliography}
\end{document}